\documentclass[twocolumn,aps,prl,showpacs,groupedaddress]{revtex4}

\usepackage{amsmath}
\usepackage{dcolumn}
\usepackage{epsfig}
\usepackage{graphicx}
\usepackage{latexsym}

\usepackage{epstopdf}

\begin{document}

\title{Correlations and Pair Formation in a Repulsively Interacting Fermi Gas}

\author{Christian Sanner, Edward J. Su, Wujie Huang, Aviv Keshet, Jonathon Gillen, and Wolfgang Ketterle}
\affiliation{MIT-Harvard Center for Ultracold Atoms, Research
Laboratory of Electronics, and Department of Physics, Massachusetts
Institute of Technology, Cambridge Massachusetts 02139, USA}

\begin{abstract}
A degenerate Fermi gas is rapidly quenched into the regime of strong
effective repulsion near a Feshbach resonance. The spin fluctuations
are monitored using speckle imaging and, contrary to several
theoretical predictions, the samples remain in the paramagnetic
phase for arbitrarily large scattering length. Over a wide range of
interaction strengths a rapid decay into bound pairs is observed
over times on the order of $10\,\hbar/E_F$, preventing the study of
equilibrium phases of strongly repulsive fermions. Our work suggests
that a Fermi gas with strong short-range repulsive interactions does
not undergo a ferromagnetic phase transition.
\end{abstract}

\maketitle

Many-body systems can often be modeled using contact interactions,
greatly simplifying the analysis while maintaining the essence of
the phenomenon to be studied. Such models are almost exactly
realized with ultracold gases due to the large ratio of the de
Broglie wavelength to the range of the interatomic forces
\cite{Ketterle2008}. For itinerant fermions with strong short-range
repulsion, textbook calculations predict a ferromagnetic phase
transition - the so-called Stoner instability \cite{Snoke2008}.

Here we investigate this system using an ultracold gas of fermionic
lithium atoms, and observe that the ferromagnetic phase transition
does not occur. A previous experimental study \cite{Jo2009}
employing a different apparatus found indirect evidence for a
ferromagnetic phase, but did not observe the expected domain
structure, possibly due to the lack of imaging resolution. Here we
address this shortcoming by analyzing density and spin density
fluctuations via speckle imaging \cite{Sanner2011}. When spin
domains of $m$ atoms form, the spin density variance will increase
by a factor of $m$ \cite{Footnote1}, even if individual domains are
not resolved. One main result of this paper is the absence of such a
significant increase which seems to exclude the possibility of a
ferromagnetic state in the studied system.


The Stoner model assumes a two-component Fermi gas with a repulsive
short-range interaction described by a single parameter, the
scattering length.  The predicted phase transition to a
ferromagnetic states requires large repulsive scattering lengths on
the order of the interatomic spacing. They can be realized only by
short-range \textit{attractive} potentials with a loosely bound
state with binding energy $\hbar^2/(m a^2)$, with $m$ being the
atomic mass and $a$ being the scattering length \cite{Footnote2}.
However, the repulsive gas is then by necessity only metastable with
respect to decay into the bound state. Many theoretical studies of a
Fermi gas with strong short-range repulsive interactions assume that
the metastable state is sufficiently long-lived
\cite{Houbiers1997,Amoruso1999,Salasnich2000,Sogo2002,
Duine2005,Zhai2009,Conduit2009,LeBlanc2009,Berdnikov2009,Cui2010,Zhang2010,Heiselberg2011}.
In recent Monte-Carlo simulations, the paired state is projected out
in the time evolution of the system \cite{Pilati2010,Chang2011}.
Theoretical studies concluded that the pairing instability is
somewhat faster than the ferromagnetic instability \cite{Pekker2011,
Sodemann2012}. The second major result of this paper is to show that
pair formation occurs indeed on a very short time scale. The
measured time constant of $10 \hbar/E_F$ (where $E_F$ is the Fermi
energy) indicates that the metastable repulsive state will never
reach equilibrium and that, even in a metastable sense, a Fermi gas
with strong short-range repulsive interactions does not exist.  The
fast pair formation could not be observed previously due to limited
time resolution \cite{Jo2009}. Instead, a much slower second phase
in the conversion of atoms to pairs was observed leading to the
wrong conclusion that the unpaired atoms have a much longer
lifetime.

\begin{figure}[tbh!]
\begin{center}
\includegraphics[width=3.5in]{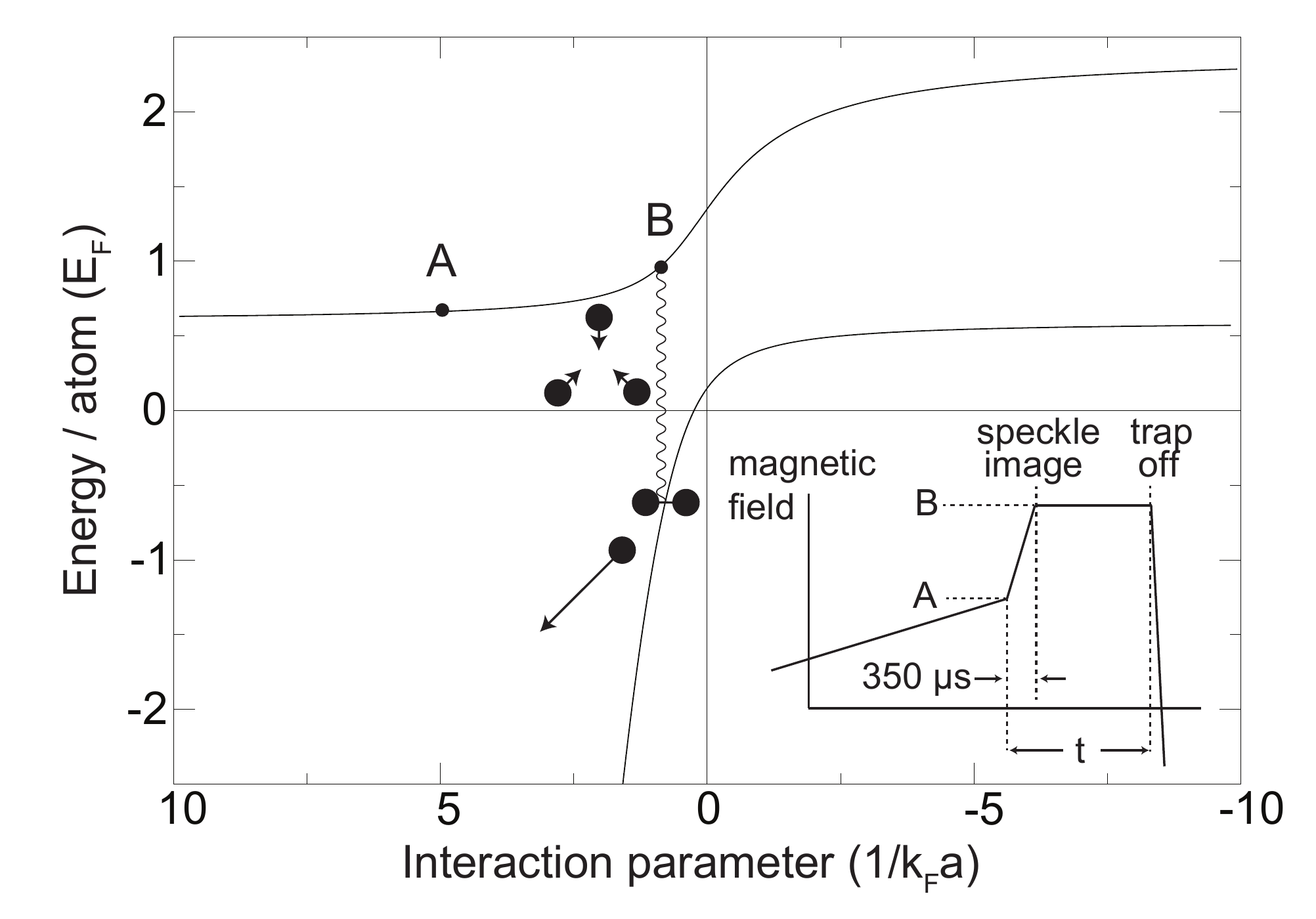}
\caption[]{Diagram showing energy levels and timing of the
experiment. The upper (repulsive) and lower (attractive) branch
energies, near a Feshbach resonance, are connected by 3-body
collisions. In our experiment, we quickly jump from a weakly
interacting Fermi gas (A) to a strongly interacting one (B) with a
rapid magnetic field change. The evolution of correlations and
domains and the molecule formation (population of the lower branch)
are studied as a function of hold time $t$. Adapted from
\cite{Pricoupenko2004}. \label{f:cartoon}}
\end{center}
\end{figure}

The experiments were carried out with typically $4.2 \times 10^{5}$
$^{6}$Li atoms in each of the two lower spin states $|1\rangle$ and
$|2\rangle$ confined in an optical dipole trap with radial and axial
trap frequencies $\omega_r=2\pi \times 100(1)$s$^{-1}$ and
$\omega_z=2\pi \times 9.06(25)$s$^{-1}$. The sample was
evaporatively cooled at a magnetic bias field $B=320 G$, identical
to the procedure described in \cite{Sanner2010}. Then the magnetic
field was slowly ramped to 730 G ($k_Fa=0.35$) in 500 ms. The
fraction of atoms being converted to molecules during the ramp was
measured (see below for method) to be below 5 \%. The temperature of
the cloud was typically $0.23(3)\,T_F$ at 527 G with a Fermi energy
of $E_F= k_B T_F= h\times6.1\,\mathrm{kHz}$. After rapidly switching
the magnetic field from 730 G to the final value in less than 350
$\mu$s, spin fluctuations were measured by speckle imaging.
Optionally an appropriate RF pulse was applied directly before
imaging to rotate the spin orientation along the measurement axis.
Due to the use of 20 cm diameter coils outside the vacuum chamber,
the inductance of the magnet coils was $330\,\mu$H  and the fast
switching was accomplished by rapidly discharging capacitors charged
to 500V.

Experimentally, spin fluctuations are measured using the technique
of speckle imaging described in Ref. \cite{Sanner2011}. For an
appropriate choice of detuning, an incident laser beam experiences a
shift of the refractive index proportional to the difference between
the local populations of the two spin states $N_1$ and $N_2$. Spin
fluctuations create spatial fluctuations in the local refractive
index and imprint a phase pattern into the incoming light, which is
then converted into an amplitude pattern during propagation. The
resulting spatial fluctuations in the probe laser intensity are used
to determine the spin fluctuations in the sample.



In Ref. \cite{Sanner2011} we prepared samples on the lower branch of
the Feshbach resonance, where positive values of $k_F a$ correspond
to a gas of weakly bound molecules. At $k_F a = 1.2$, we observed a
sixfold suppression of spin fluctuations  and a fourfold enhancement
of density fluctuations. Typical fluctuations in the speckle images
of a non-interacting Fermi gas at $T = 0.23 T_F$ amount to 5 \% of
the average optical signal per pixel, corresponding to about 50 \%
of Poissonian fluctuations. Those fluctuations are modified by
factors between 0.2 and 1.6 due to pairing and interactions.

In this study, on the upper branch of the Feshbach resonance, the
situation is reversed. For unbound atoms, as the interaction
strength increases, the two spin components should develop stronger
and stronger anticorrelations and enhanced spin fluctuations.
Previous experimental work \cite{Jo2009} and several theoretical
studies \cite{Sogo2002, Duine2005, Conduit2009, LeBlanc2009,
Berdnikov2009, Heiselberg2011} predicted a phase transition to a
ferromagnetic state where the magnetic susceptibility and therefore
the spin fluctuations diverge. Recent Monte Carlo simulations
\cite{Pilati2010} predict such a divergence around $k_F a = 0.83$.
We therefore expected an increase of spin fluctuations by one or
several orders of magnitude, related to the size of magnetic
domains.

\begin{figure}[tbh!]
\begin{center}
\includegraphics[width=3.5in]{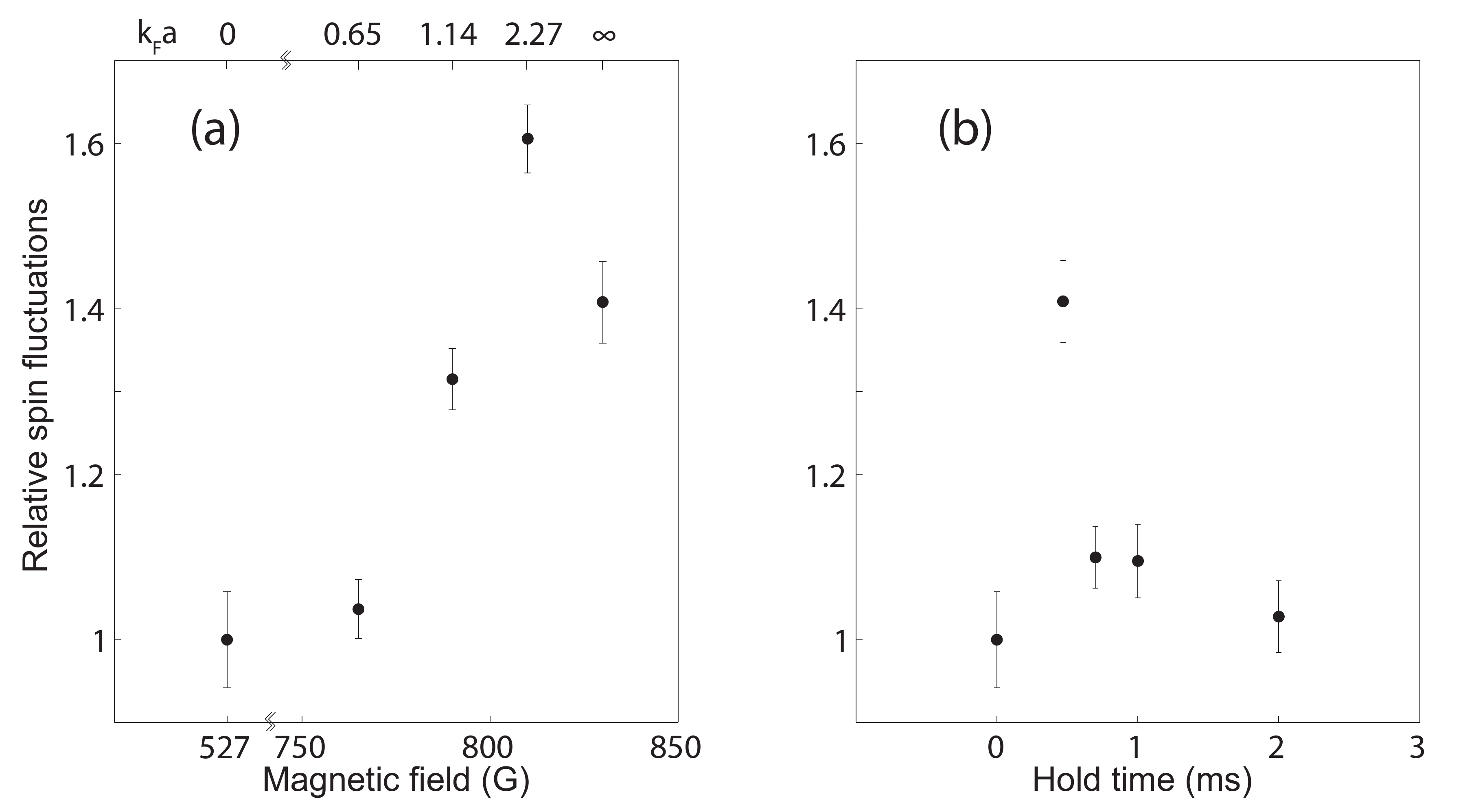}
\caption[]{Spin fluctuations (a) after 350 $\mu$s as a function of
magnetic field and (b) on resonance as a function of hold time
scaled to the value measured at 527G. Even at strong repulsive
interactions, the measured spin fluctuations are barely enhanced,
indicating only short-range correlations and no domain formation.
The spin fluctuations were determined for square bins of 2.6 $\mu$m,
each containing on average 1000 atoms per spin
state.\label{f:noise}}
\end{center}
\end{figure}

Figure \ref{f:noise} shows the observed spin fluctuations
enhancement compared to the non-interacting cloud at 527 G. The
variance enhancement factor reaches its maximum value of 1.6
immediately after the quench, decreasing during the 2 ms afterward.
The absence of a dramatic increase shoes that no domains form and
that the sample remains in the paramagnetic phase throughout.
Similar observations were made for a wide range of interaction
strengths and wait times. Note that first-order perturbation theory
\cite{Recati2011} predicts an increase of the susceptibility by a
factor of 1.5 at $k_F a = 0.5$ and by a factor of 2 at $kFa = 0.8$
(i.e. no dramatic increase for $k_F a < 1$). Therefore, our data
shows no evidence for the Fermi gas approaching the Stoner
instability.

Before we can fully interpret these findings, we have to take into
account the decay of the atomic sample on the upper branch of the
Feshbach resonance into bound pairs. We characterize the pair
formation by comparing the total number of atoms and molecules
$N_\mathrm{a}$+2$N_{\mathrm{mol}}$ (determined by taking an
absorption image after ballistic expansion at high magnetic field
where molecules and atoms have the same absorption resonance) to the
number of free atoms (determined by rapidly sweeping the magnetic
field to 5G before releasing the atoms and imaging the cloud,
converting pairs into deeply bound molecules that are completely
shifted out of resonance) \cite{Jochim2003}.

\begin{figure}[tbh!]
\begin {center}
\includegraphics[width=3.5in]{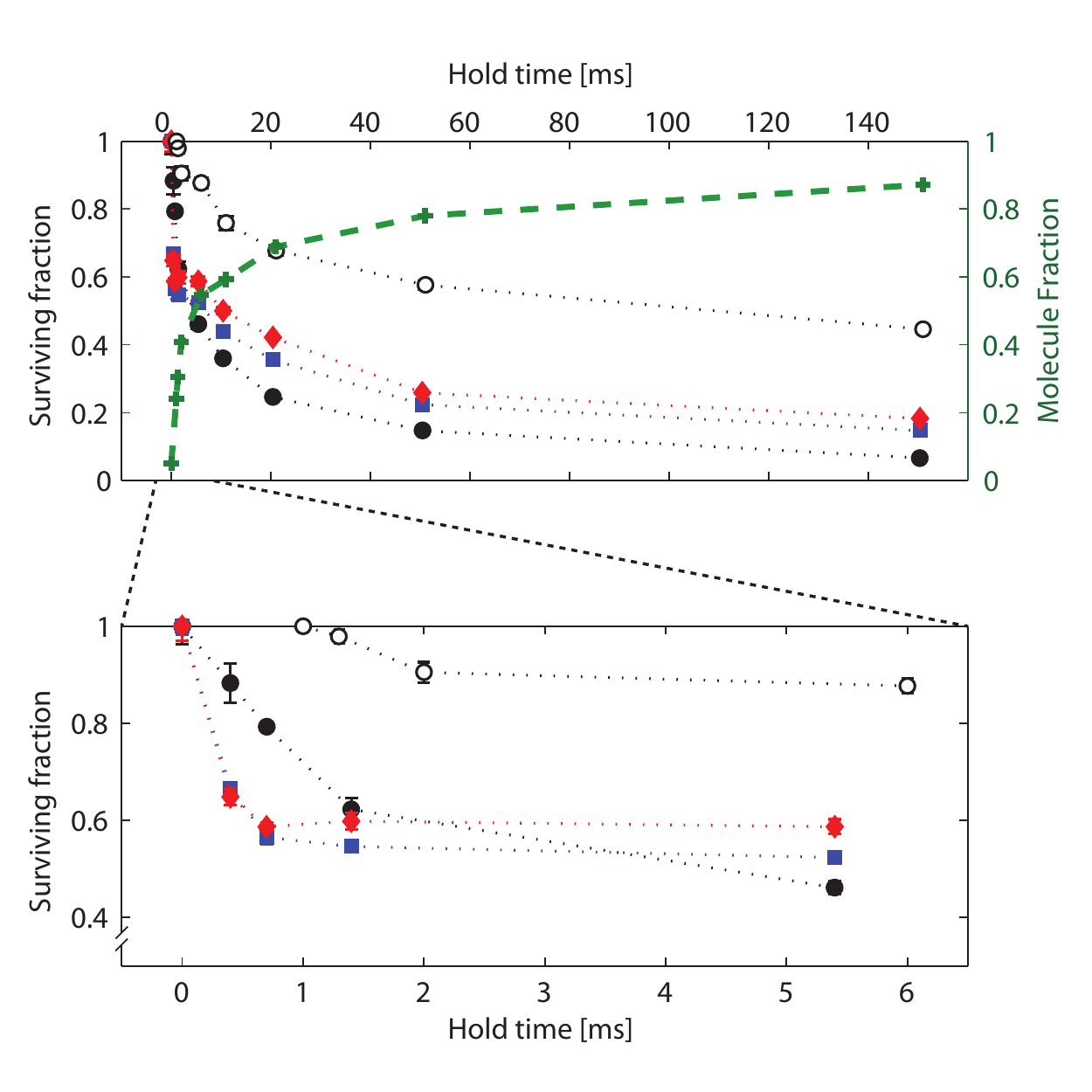}
\caption[]{Characterization of molecule formation at short and long
hold times, and at different values of the interaction strength. The
closed symbols, circles (black) at 790G with $k_F a = 1.14$, squares
(blue) at 810G with $k_F a = 2.27$ and diamonds (red) at 818G with
$k_F a = 3.5$ represent the normalized number of free atoms, the
open symbols the total number of atoms including those bound in
Feshbach molecules (open circles at 790G with $k_F a = 1.14$). The
crosses (green) show the molecule fraction. The characteristic time
scale is set by the Fermi time $\hbar / E_F = 43 \mu$s, calculated
with a cloud averaged Fermi energy. \label{f:molecules}}
\end{center}
\end{figure}

The time evolution of the molecule production (Fig.
\ref{f:molecules}) shows two regimes of distinct behavior. For times
less than 1 ms, we observe a considerable number of atoms converted
into molecules, while the total number
$N_\mathrm{a}$+2$N_{\mathrm{mol}}$ remains constant. The initial
drop in atom number becomes larger as we increase the final magnetic
field, and saturates at around 50 \% near the Feshbach resonance.

We attribute this fast initial decay in atom number to recombination
\cite{Chin2010,Conduit2011} into the weakly bound molecular state.
We obtain an atom loss rate $\dot{N}_a/N_a=250\,\mathrm{s}^{-1}$ at
790G in the first 1ms after the magnetic field switch. Assuming a
three-body process we estimate the rate coefficient $L_3$ at this
field to be $3.9\times10^{-22}\,\mathrm{cm}^6\,\mathrm{s}^{-1}$,
though the interaction is already sufficiently strong for many-body
effects to be significant. For stronger interactions, about 30\% of
atom loss occurs already during the relevant 100 $\mu$s of ramping
through the strongly interacting region, indicating a lower bound of
around $3\times10^3\,\mathrm{s}^{-1}$ for the loss rate which is
13\% of the inverse Fermi time $E_F/\hbar$, calculated with a cloud
averaged Fermi energy.

After the first millisecond, the molecule formation rate slows down,
by an order of magnitude at a magnetic field of 790G (and even more
dramatically at higher fields) when it reaches about 50 \%. It seems
likely that the molecule fraction has reached a quasi-equilibrium
value at the local temperature, which is larger than the initial
temperature due to local heating accompanying the molecule
formation. Ref. \cite{CChin2004} presents a simple model for the
equilibrium between atoms and molecules (ignoring strong
interactions). For phase space densities around unity and close to
resonance, the predicted molecule fraction is 0.5, in good agreement
with our observations \cite{Footnote3}.


For longer time scales (hundred milliseconds) we observe a steady
increase of the molecule fraction to 90 \% for the longest hold
time.  This occurs due to continuous evaporation which cools down
the system and shifts the atom-molecule equilibrium towards high
molecule fractions.  During the same time scale, a slow loss in both
atom number and total number is observed caused by inelastic
collisions (vibrational relaxation of molecules) and evaporation
loss.

Is the rapid conversion into molecules necessarily faster than the
evolution of ferromagnetic domains?  Our answer is tentatively yes.
First, for strong interactions with $k_F a$ around 1, one expects
both instabilities (pair formation and Stoner instability) to have
rates which scale with the Fermi energy $E_F$ and therefore with
$n^{2/3}$. Therefore, one cannot change the competition between the
instabilities by working at higher or lower densities. According to
Ref. \cite{Pekker2011} the fastest unstable modes for domain
formation have a wavevector $q \approx k_F / 2$ and grow at a rate
of up to $E_F/4\hbar$ when the cloud is quenched sufficiently far
beyond the critical interaction strength.  Unstable modes with such
wavevectors will develop ``domains'' of half a wavelength or size
$\xi = \pi/q = 2\pi/k_F$ containing 5 atoms per spin state in a
volume $\xi^3$. This rate is comparable to the observed conversion
rates into pairs of $0.13 E_F$.  Therefore, at best, ``domains'' of
a few particles could form, but before they can grow further and
prevent the formation of pairs (in a fully polarized state), rapid
pair formation takes over and populates the lower branch of the
Feshbach resonance.  Based on our observations and these arguments,
it seems that it is not possible to realize ferromagnetism with
strong short range interaction, and therefore the basic Stoner model
cannot be realized in nature.

One possibility to suppress pair formation is provided by narrow
Feshbach resonances.  Here the pairs have dominantly closed channel
character and therefore a much smaller overlap matrix element with
the free atoms. However, narrow Feshbach resonances are
characterized by a long effective range and do not realize the
Stoner model which assumes short-range interactions. Other
interesting topics for future research on ferromagnetism and pair
formation include the effects of dimensionality \cite{Drummond2011,
Pekker2012}, spin imbalance \cite{Liu2010a,Dong2010}, mass imbalance
\cite{Keyserlingk2011}, lattice and band structure
\cite{Chang2010,Carleo2011}.

We now discuss whether ferromagnetism is possible \emph{after} atoms
and molecules have rapidly established local equilibrium. In other
words, starting at $T=0$, one could heat up the fully paired and
superfluid system and create a gas of atomic quasiparticles which
are similar to free atoms with repulsive interactions. Density and
temperature of the atoms are now coupled.  It is likely that such a
state is realized in our experiments after a few ms following the
quench, until evaporative cooling converts the system into a
molecular condensate over $\approx$ 100 ms. The possibility that
such a quasiparticle gas could become ferromagnetic has not been
discussed in the literature. Our experiments do not reveal any major
increase in spin fluctuations which seems to exclude a ferromagnetic
state. In the simplest picture, we could regard the atomic
quasiparticles as free atoms, and then apply the Stoner model to
them. Ferromagnetic domain formation is analogous to phase
separation between the two spin components \cite{Jo2009}. Since
dimers interact equally with the two spin components, one might
expect that even a noticeable dimer fraction should not suppress the
tendency of the atomic gas to form domains. Therefore, in a simple
model, one may neglect dimer-atom interactions.

If the Stoner model applies to this quasiparticle gas, the next
question is whether the temperature is low enough for the
ferromagnetic phase transition.  Available theoretical treatments do
not predict an exact maximum transition temperature to the
ferromagnetic state and obtain an unphysical divergence for large
scattering lengths. Since the only energy scale is the Fermi
temperature, one would expect a transition temperature which is a
fraction of the Fermi temperature \cite{Liu2010b}, higher or around
the temperature scale probed in our experiments. However, even above
the transition temperature, the susceptibility is enhanced. A simple
Weiss mean field or Stoner model leads to the generic form of the
susceptibility $\chi(T) = \chi_0(T)/(1-w\chi_0(T))$, where
$\chi_0(T)$ is the Pauli susceptibility of the non-interacting gas
and $w$ the interaction parameter. This formula predicts a two-fold
increase in the susceptibility even 50 \% above the transition
temperature, which is well within the sensitivity of our
measurements.

Therefore, our experiment can rule out ferromagnetism for
temperatures even slightly lower than the experimental temperatures.
Temperatures are very difficult to measure in a transient way for a
dynamic system which may not be in full equilibrium.  For example,
cloud thermometry requires full equilibration and lifetimes much
longer than the longest trap period.  We attempted to measure the
temperature after the hold time near the Feshbach resonance by
quickly switching the magnetic field to weak interactions at 527 G
and then performing noise thermometry using speckle imaging
\cite{Sanner2011}. We measure column-integrated fluctuations that
are 0.61(8) of the Poisson value which implies an effective
temperature well below $T_F$, around 0.33(7) $T_F$, not much higher
than our initial temperature of 0.23 $T_F$. Although the cloud is
not in full equilibrium, an effective local temperature can still be
obtained from noise thermometry.

Alternatively, we can estimate the temperature increase from the
heat released by pair formation. A simple model
\cite{Supplement2011} accounting for the relevant energy
contributions predicts for $k_F a = 1$ that
%
%
molecule fractions of higher than 20$\%$ result in a final
temperature above 0.4$T_F$, an estimate which is higher than the
measurement reported above. One may hope that closer to resonance
many-body effects lower the released energy, however as we show in
the supplement (Fig. 1 of \cite{Supplement2011}) this is not
necessarily the case due to the repulsive interaction energy.

Our experiment has not shown any evidence for a possible
ferromagnetic phase in an atomic gas in ``chemical'' equilibrium
with dimers.  This implies one of the following possibilities.  (i)
This gas can be described by a simple Hamiltoninan with strong short
range repulsion. However, this Hamiltonian does not lead to
ferromagnetism.  This would be in conflict with the results of
recent quantum Monte-Carlo simulations \cite{Pilati2010,Chang2011}
and second order perturbation theory \cite{Duine2005}, and in
agreement with conclusions based on Tan relations \cite{Barth2011}.
(ii) The temperature of the gas was too high to observe
ferromagnetism.  This would then imply a critical temperature around
or below $0.2 T/T_F$, lower than generally assumed. (iii) The
quasiparticles cannot be described by the simple model of an atomic
gas with short-range repulsive interactions due to their
interactions with the paired fraction.

A previous experiment \cite{Jo2009} reported evidence for
ferromagnetism by presenting non-monotonic behavior of atom loss
rate, kinetic energy and cloud size when approaching the Feshbach
resonance, in agreement with predictions based on the Stoner model.
Our measurements confirm that the properties of the gas strongly
change near $k_F a=1$. Similar to \cite{Jo2009}, we observe features
in kinetic and release energy measurements near the resonance (see
Supplemental Information \cite{Supplement2011}). However, the
behavior is more complex than that captured by simple models. The
atomic fraction decays non-exponentially (see Fig.
\ref{f:molecules}), and therefore an extracted decay time will
depend on the details of the measurement such as time resolution.
Ref. \cite{Jo2009} found a maximum of the loss rate of
$200\,\mathrm{s}^{-1}$ for a Fermi energy of 28 kHz. Our lower bound
of the decay rate of $3 \times 10^3\,\mathrm{s}^{-1}$ is 15 times
faster at a five times smaller Fermi energy. Our more detailed study
rules out that Ref. \cite{Jo2009} has observed ferromagnetic
behavior.

Our conclusion is that an ultracold gas with strong short range
repulsive interactions near a Feshbach resonance remains in the
paramagnetic phase. The fast formation of molecules and the
accompanying heating makes it impossible to study such a gas in
equilibrium, confirming predictions of a rapid conversion of the
atomic gas to pairs \cite{Pekker2011, Zhang2011}. Therefore, it
appears that the widely used Stoner model cannot be realized in
Nature in its simplest form since the neglected competition with
pairing is crucial.

This work was supported by NSF and ONR, AFOSR MURI, and under ARO
grant no. W911NF-07-1-0493 with funds from the DARPA Optical Lattice
Emulator program. We are thankful to Eugene Demler, David Pekker,
Boris Svistunov, Nikolay Prokof'ev, and Wilhelm Zwerger for valuable
discussions and to David Weld for critical reading of the
manuscript.

\bibliographystyle{aip}

\begin{thebibliography}{10}
\bibitem{Footnote1}
This is illustrated by a simplified model assuming {P}oissonian fluctuations in
  a given probe volume within the atom sample. {W}ith on average ${N}$ atoms in
  this volume one would measure a standard deviation in the atom number of
  $\sqrt{N}$. {H}owever, if the atoms formed clusters each made of $m$ atoms
  the standard deviation of the number of clusters would be $\sqrt{N/m}$,
  leading to a variance in atom number of $(m\sqrt{N/m})^2=m {N} $.
\bibitem{Footnote2}
Purely repulsive potentials have an extent of at least the scattering length
  and are therefore not short-range for $k_{F} a \geq 1$.
\bibitem{Footnote3}
Note that the drop in molecule formation after 1 ms cannot be explained by the
  drop in atomic density by a factor of two due to conversion into molecules
  and to an increase in the size of the sample due to the increased repulsive
  interactions.
\end{thebibliography}

\end{document}